\documentclass[reprint, nocopyrightspace]{sigplanconf}

\usepackage{amsmath}
\usepackage{amssymb}
\usepackage{graphics}
\usepackage{graphicx}
\usepackage[scriptsize]{subfigure}

\usepackage{caption} 
\usepackage{listings}
\usepackage{color}
\usepackage{xcolor}
\usepackage{booktabs}
\usepackage{amsmath}

\definecolor{mygreen}{rgb}{0,0.6,0}
\definecolor{mygray}{rgb}{0.5,0.5,0.5}
\definecolor{mymauve}{rgb}{0.58,0,0.82}

\begin{document}

\conferenceinfo{}{} 
\copyrightyear{2016} 
\copyrightdata{} 
\doi{}
%
\title{2D Image Convolution using Three Parallel Programming Models on the Xeon Phi}
%
%
%

\authorinfo{Ashkan Tousimojarad}
           {University of Glasgow}
           {ashkan.tousi@gmail.com}
\authorinfo{Wim Vanderbauwhede}
           {University of Glasgow}
           {wim.vanderbauwhede@glasgow.ac.uk}
\authorinfo{W Paul Cockshott}
		{University of Glasgow}
		{wpc@dcs.gla.ac.uk}

\maketitle

\begin{abstract}
Image convolution is widely used for sharpening, blurring and edge detection. In this paper, we review two common algorithms for convolving a 2D image by a separable kernel (filter). After optimising the naive codes using loop unrolling and SIMD vectorisation, we choose the algorithm with better performance as the baseline for parallelisation. We then compare the parallel performance of the optimised code using OpenMP, OpenCL and GPRM implementations on the Intel Xeon Phi. 
We also measure the effects of optimisation techniques and demonstrate how they affect both sequential and parallel performance.
Apart from comparing the code complexity as well as the performance of the chosen parallel programming models, we investigate the impact of a parallelisation technique, task agglomeration in GPRM. 
\end{abstract}


%

\section{Introduction}
%
%
%
%
Throughput computing applications demand for fast response time while dealing with a large amount of data. Image convolution is one of such throughput computing applications. Convolution of an image by a matrix of real numbers can be used to sharpen or smooth an image, depending on the matrix used. If $A$ is an image and $K$ is a convolution matrix, then $B$, the convolved image is calculated as:

\begin{equation}‎\label{eq:convformula}
B_{y , x} = \sum\limits_{i} \sum\limits_{j} A_{y+i , x+j}K_{i , j}
\end{equation}‎

If $\textbf{k}$ is a convolution vector, then the corresponding matrix $K$ is such that $K_{i , j} = \textbf{k}_i\textbf{k}_j$

A separable convolution kernel is a vector of real numbers that can be decomposed into horizontal and vertical projections and hence can be applied independently to the rows and columns of the spatial domain to provide filtering \cite{smith2013digital}. It is a specialisation of the more general convolution, but is algorithmically more efficient to implement. 

The image convolution algorithms are taken from the real code used in a stereo matching algorithm. Image convolution and scaling take up most of the cycles in the stereo matching algorithm. For all tests, separable kernels of width 5 are used. 
The algorithm uses 3 colour planes and is heavily memory-fetch bound.

McCool et al. \cite{mccool2012structured} list three desired features for the parallel programming models that intend to enable parallelism: I) Performance, II) Productivity and III) Portability. It should be possible to predict good performance, tune it, and scale it to larger systems. Productivity is not only about expressiveness and composability, but also about maintainability. Supporting a range of targets and operating systems is another desirable property, known as portability. We have considered these three features in choosing the parallel models for this study and we will refer to them throughout the paper.

We aim to explore a number of optimisation and parallelisation techniques for enhancing the performance of 2D image convolution on a modern manycore architecture. For this purpose, we have chosen three parallel programming models from different domains and for different reasons: 
OpenMP,  the de-facto standard for programming shared memory architectures; OpenCL, known for being portable across multiple platforms; and finally GPRM,  a pure task-based programming framework. It has been reported that GPRM provides superior performance compared to OpenMP on manycore architectures \cite{tousimojarad2014parallel} \cite{tousimojaradSteal}. 

This paper is structured as follows: we start by introducing the Xeon Phi and the three parallel models in sections \ref{conv:xeonphi} and \ref{conv:parmodels}, followed by the experimental setup in section \ref{conv:setup}. In section \ref{convImplement}, we review the two algorithms used to solve the problem: single-pass and two-pass algorithms. In this section, we also list different optimisation techniques to convert a naive code to an optimised one. We then discuss the implementation details of the parallel two-pass algorithm in each programming model. Results of parallelising the two-pass algorithm is presented in section \ref{conv:2-passResults}. In section \ref{conv:nocpy1-passResults}, another version of the single-pass algorithm is considered. We will show that although, still the sequential two-pass algorithm outperforms the sequential single-pass algorithm, the parallel performance of the modified single-pass algorithm is better. Finally, we review a few of related research studies and the Conclusion section summarises our attempts in exploring the behaviour of the two algorithms, pros and cons of the studied parallel models in solving this problem on a manycore architecture as well as the effect of optimisation and parallelisation techniques applied to improve the performance. 

\section{Hardware Platform: Intel Xeon Phi}\label{conv:xeonphi}

Contemporary applications have increased the trend of integrating a large number of
processing cores in order to meet their performance goals. Most of these
applications need single-chip implementation to satisfy their size and power consumption requirements. Multicore and manycore processors have emerged as promising architectures to benefit from increasing transistor numbers.

The Intel Xeon Phi coprocessor 5110P used in this study is an SMP (Symmetric Multiprocessor) on-a-chip which is connected to a host Xeon processor via a PCI Express bus interface. The Intel Many Integrated Core (MIC) architecture used by the Intel Xeon Phi coprocessors gives developers the advantage of using standard, existing programming tools and methods. Our Xeon Phi comprises 60 cores (240 logical cores) connected by a bidirectional ring interconnect.

The Xeon Phi coprocessor provides four hardware threads sharing the same physical core and its cache subsystem in order to hide the latency inherent in in-order execution. As a result, the use of at least two threads per core is almost always beneficial \cite{jeffers2013intel}. The Xeon Phi has eight memory controllers supporting 2 GDDR5 memory channels each. The clock speed of the cores is 1.053GHz. Each core has an associated 512KB L2 cache. Data and instruction L1 caches of 32KB are also integrated on each core. Another important feature of the Xeon Phi is that each core includes a SIMD 512-bit wide VPU (Vector Processing Unit).
The VPU can be used to process 16 single-precision or 8 double-precision elements per clock cycle.

\section{Parallel Programming Models}\label{conv:parmodels}
The design of manycore processors is strongly driven by demands for greater performance at reasonable cost. To make effective use of the available parallelism in such systems, the parallel programming model is of great importance. There are several parallel programming models, runtime libraries, and APIs that help developers to move from sequential to parallel programming. For the purposes of this study, we have chosen three parallel programming models that the Xeon Phi supports: OpenMP, OpenCL and GPRM.

\subsection{OpenMP}

OpenMP is the de-facto standard for shared memory programming, and is based on a set of compiler directives or pragmas, combined with a programming API to specify parallel regions, data scope, synchronisation, and so on. It also supports runtime configuration through the use of runtime environment variables, e.g. \texttt{OMP\_NUM\_THREADS} to specify the number of threads at runtime. OpenMP is a portable parallel programming approach and is supported on C, C++, and Fortran. It has been historically used for loop-level and regular parallelism through its compiler directives. Since the release of OpenMP 3.0, OpenMP also supports task parallelism \cite{ayguade2009design}. It is now widely used in both task and data parallel scenarios. 

The Intel OpenMP runtime library (as opposed to the GNU implementation) allocates a task list per thread for every OpenMP team. Whenever a thread creates a task that cannot not be executed immediately, that task is placed into the thread's deque (double-ended queue). A random stealing strategy balances the load \cite{clet2014evaluation}.

Since OpenMP is a language enhancement, every new construct requires compiler support. Therefore, its functionality is not as extensive as library-based models. Moreover, race condition is a serious problem in OpenMP. Although it provides the user with a high level of abstraction, the onus is still on the programmer to avoid race condition.

\subsection{OpenCL}
OpenCL \cite{stone2010opencl} is an open standard for heterogeneous architectures. One of the main OpenCL's objectives is to increase portability across GPUs, multicore processors, and OS software via its abstract memory and execution model; however its performance is not always portable across different platforms. It has been suggested to consider the architectural specifics in the algorithm design, in order to address its performance portability issue. The use of auto-tuning heuristics could also improve the performance \cite{du2012cuda}.

Although OpenCL is mostly compared against Nvidia's CUDA \cite{sanders2010cuda}, we do not aim to cover discussions about GPUs in this work. The reason why OpenCL is listed here is firstly because it allows for sharing of workload between host and device with the same program, hence increasing portability and productivity, and secondly because the studied system, the Xeon Phi,  supports the OpenCL programming model. 

\subsection{GPRM}
The Glasgow Parallel Reduction Machine (GPRM) \cite{tousimojaradSteal} provides a task-based approach to manycore programming by structuring programs into \emph{task code}, written as C++ classes, and \emph{communication code}, written in GPC, a restricted subset of C++. The communication code describes how the \emph{task}s interact. Compiling a GPRM program results in an Intermediate Representation (IR) including useful
information about tasks, their dependencies as well as the initial task-to-thread mapping information. GPRM is not as well-known as the other two approaches, but since it specifically targets manycore architectures and  has shown good performance compared to OpenMP and other approaches on the Xeon Phi, we have used it to see how well a ``pure'' task-based model performs for this benchmark. 

The aim of GPRM is to abstract away all the details of threads, promoting the idea that \emph{task}s are the actual computations, and the threads are only their substrates. What this means in practice is that the runtime system automatically creates as many threads as the number of cores. The programmer only decides about the tasks' details, i.e. the number of tasks (using a cutoff), the task code and the communication pattern between the tasks.

GPRM promises a good performance by combining compile-time (source to intermediate representation) and runtime (stealing) techniques. In other words, compile-time decisions form the initial task distribution and the runtime system adjusts dynamically. As a result, some threads can be asleep during the execution, which means that the number of active threads are tuned automatically. An implication of this model is that if the scheduling and/or communication overhead is significant, one should create fewer tasks than the number of cores.

\section{Experimental Setup}\label{conv:setup}
OpenMP and GPRM programs are executed natively on the MIC. The Intel compiler \texttt{icpc (ICC) 14.0.2} is used with the \texttt{-mmic}, \texttt{-O2}, \texttt{-no-offload} and \texttt{-ansi-alias} flags for compiling the programs for native execution on the Xeon Phi. The OpenMP programs should be compiled with the \texttt{-openmp} flag and the \texttt{libiomp5.so} library is required. Unlike OpenMP, for the GPRM framework there is no shared library to be copied to the MIC. 

We also used OpenCL (Intel OpenCL for MIC v1.2) with \texttt{-cl-mad-enable} and \texttt{-cl-fast-relaxed-math}. As we want to compare the compute performance on the MIC, the time to transfer the image between the host and device in the offload mode is not taken into account. 

Threads and tasks are two completely different concepts. However, in a pure task-based model such as GPRM, concurrency is only controlled by tasks. Suppose that we want to parallelise a simple loop on N elements. In OpenMP, using 100 threads simply means that each thread gets N/100 of the workload (if the static scheduling is chosen). In GPRM for every program, the number of tasks should be specified. For a loop, each chunk corresponds to a task, therefore 100 tasks (i.e. cutoff=100) on the Xeon Phi means that each thread gets N/100 of the workload. If we choose a cutoff=480, since there are 240 threads on the Xeon Phi, each thread gets 2 tasks.
GPRM speedup is therefore achieved by changing the number of tasks rather than the number of threads. In the case of the Xeon Phi, the number of threads is set to 240 by the GPRM runtime system. 

All speedup ratios are computed against the running time of the sequential code implemented in C++, which means they are directly proportional to the absolute running times. 
There will be 3 planes per image and the benchmark will be run for 1000 times in order to measure a precise running time on the MIC. Therefore, the time for each image should be considered as $runningtime/1000$.

We have considered a Gaussian separable 5$\times$5 kernel and a range of square images from 1152$\times$1152 to 8748$\times$8748 for the purposes of this study. We refer two implementations of the convolution algorithm as single-pass and two-pass algorithms (implementations). The single-pass algorithm is the general algorithm used for convolution, having a nested loop over the kernel, therefore comprised of four loops to compute the convolution. The two-pass algorithm is specific to separable kernels, and uses a horizontal 1D convolution pass followed by a vertical 1D convolution pass to convolve the image.

Authors in \cite{chimehcompiling} have identified that the peak performance for the two-pass algorithm occurs at 100 threads. Our initial experiments has verified that considering the range of images from 1152$\times$1152 to 8748$\times$8748, 100 could be our magic number for both OpenMP (optimal number of threads) and GPRM (optimal number of tasks) models. It is worth stating that because the convolution time for each image is too short, the communication overhead becomes significant, and using all of the available resources in the Xeon Phi is not advantageous. This is not the case for OpenCL, and using less than the available resources in terms of compute units and processing items does not improve the performance.

\section{Convolution Implementations}\label{convImplement}
The convolution algorithm used in the real world application only works at the central part of the image that is in sight of multiple cameras, and what happens at the far edges are ignored. Therefore, we can safely ignore the complications at the edges and start the convolution from the pixel for which the kernel can access the required neighbours, i.e. pixel (2,2).

\subsection{Single-Pass and Two-pass Algorithms}
In order to solve the 2D convolution problem, the simplest approach is to loop over all the image pixels and all the kernel elements in one go. This algorithm in referred as the single-pass algorithm in this study. It uses 4 nested loops, the 2 outer loops on the rows and columns and the image and the 2 inner loops on the rows and columns of the kernel. For a 5$\times$5 kernel, according to Eq. \ref{eq:convformula}, it needs 25 multiply-accumulate operations for each pixel.

An alternative comes into play when a separable kernel is used: the two-pass algorithm. As stated in the Introduction, a separable kernel can be decomposed into horizontal and vertical projections and hence can be applied independently to the rows and columns of the spatial domain. For a 5$\times$5 kernel, it reduces the number of multiplications per pixel to 10.

From the algorithmic point of view, the two-pass convolution algorithm should always be the preferred one if the kernel is separable. It has $O(n)$ time complexity, while the complexity of single-pass algorithm is $O(n^{2})$, where \emph{n} is the kernel width.

\subsection{From Naive to Parallelised Optimised Code}
We have implicitly mentioned the two-pass algorithm as an optimisation for the single-pass algorithm. We discuss in this section and section \ref{conv:nocpy1-passResults} that one should be careful about parallel performance prediction based on the sequential runtime of algorithms. For the purposes of this section we have chosen the largest 3 images of our 6 image test cases. 

There is a number of other optimisations at different levels that could be considered for image convolution. Nevertheless, we do not consider our final optimised code as a Ninja code, i.e. the best optimised solution. The optimisations listed here can be achieved with a little programming effort. The resulting optimised code, by definition should have performance comparable to a Ninja implementation, with a little effort of algorithmically improving the naive code (i.e. the compiler-generated code), or by using the latest compiler technology for parallelisation and vectorisation \cite{satish2015can}. It has also been reported that the difference between a sequential optimised (similar to our single-pass optimised code) and a sequential Ninja implementation for a 2D convolution algorithm on the Xeon Phi is around 1\% \cite{petersen2013measuring}. 

Another point to make is that in this study, we are also concerned about parallelisation techniques. An optimised sequential algorithm that utilises the vector units efficiently is important as the baseline for parallelisation, and that is why we will apply the following optimisations, but the other aspect of this study is to identify the pros and cons of parallel programming models in parallelising the optimised code.  

Here, we consider the single-pass algorithm with 4 nested loops as the naive code. It is important to note that since this algorithm convolves image array A to B, at the end of the algorithm we copy back B to A to have the original image convolved. To make sure that the naive code does not utilise automatic vectorisation, the code should be compiled with the \texttt{-no-vec} flag (although, the Intel compiler failed to auto-vectorise our naive code).  

\paragraph{Opt-1: Single-pass, Unrolled}
The first optimisation is loop unrolling. An average (among the 3 images) benefit of 2.5$\times$ can be obtained by hand unrolling the nested loop over kernel into 25 multiplications. At this stage we change the statement in Eq. \ref{eq:full} inside the kernel nested loop into 25 additions in the form of Eq. \ref{eq:unrolled}.

\begin{equation}‎ \label{eq:full}
B[pId][i][j] += A[pId][i+kx-2][j+ky-2] *K[kx][ky];
\end{equation}‎

\begin{equation}‎ \label{eq:unrolled}
\begin{split}
B[pId][i][j] =  A[pId][i-2][j-2] * K[0][0] +\\
A[pId][i-2][j-1] * K[0][1] + ... + \\
A[pId][i][j] * K[2][2] + ... + \\
A[pId][i+2][j+2] * K[4][4]; 
\end{split}
\end{equation}‎

\paragraph{Opt-2: Single-pass, Unrolled, Vectorised}
After unrolling the kernel loops, only 2 out of 4 initial loops remain. Utilising the compiler technology, we can enforce inner loop vectorisation using \texttt{\#pragma simd}, which if memory alias or dependence analysis fails, gives hint to the compiler that the loop is safe to be vectorised \footnote{This always requires extra care, as enforcing SIMD vectorisation while there is vector dependence results in incorrect results}. Vectorisation after unrolling gave us an average speedup of 22$\times$ over the baseline. 

\paragraph{Opt-3: Two-pass, Unrolled}
The third optimisation is an algorithmic change due to the fact that the kernel is separable, hence instead of 25 multiplication for each pixel (as a result of a 5$\times$5 kernel nested loop), we can use a horizontal 1D convolution followed by a 1D vertical convolution. Therefore, the number of multiplications for each pixel becomes $5+5=10$. This optimisation is first combined with loop unrolling. Each of the two loops to be unrolled in this case (one in the horizontal pass and the other in the vertical pass) has the size of 5. An average speedup of 5.5$\times$ over the baseline can be obtained at this stage.

\paragraph{Opt-4: Two-pass, Unrolled, Vectorised}
Repeating the second optimisation on the inner loops of the two-pass algorithm (i.e. those over the image columns), we can now get an average of 47.1$\times$ performance gain over the baseline, and we have just optimised the sequential code so far.

\paragraph{Par-1: Single-pass, Unrolled, 100 OpenMP Threads}
OpenMP provides the simplest way of parallelising the outer loop of the single-pass algorithm. We obtained the an average of 191.1$\times$ speedup over the baseline.

\paragraph{Par-2: Single-pass, Unrolled, Vectorised, 100 OpenMP Threads}
On top of the previous optimisation, similar to ``Opt-2'', we have enforced vectorisation on the inner loops over the image columns (for both convolution computation and the copy-back operation). Apart form that, the outer loops over the image rows are parallelised using \texttt{\#pragma omp parallel for}. An average performance gain of 1268.8$\times$ over the baseline has been achieved. 

\paragraph{Par-3: Two-pass, Unrolled, 100 OpenMP Threads}
Parallelised version of the two-pass algorithm provides an average of 393.7$\times$ speedup over the baseline. This is almost 2.1$\times$ the speedup of the competitive algorithm in ``Par-1''.

\paragraph{Par-4: Two-pass, Unrolled, Vectorised, 100 OpenMP Threads}
The best parallelised vectorised approach has the average speedup of 1611.7$\times$. This is only 1.3$\times$ the speedup of the competitive algorithm in ``Par-2''. This shows that the single-pass algorithm can benefit more from vectorisation when parallelised. This is an important finding and we will see in section \ref{conv:nocpy1-passResults} that it helps another version of the single-pass algorithm (without copy-back) to outperform the two-pass algorithm with 100 threads. 

\begin{figure}[h]
\begin{centering}
	\includegraphics[width=0.49\textwidth]{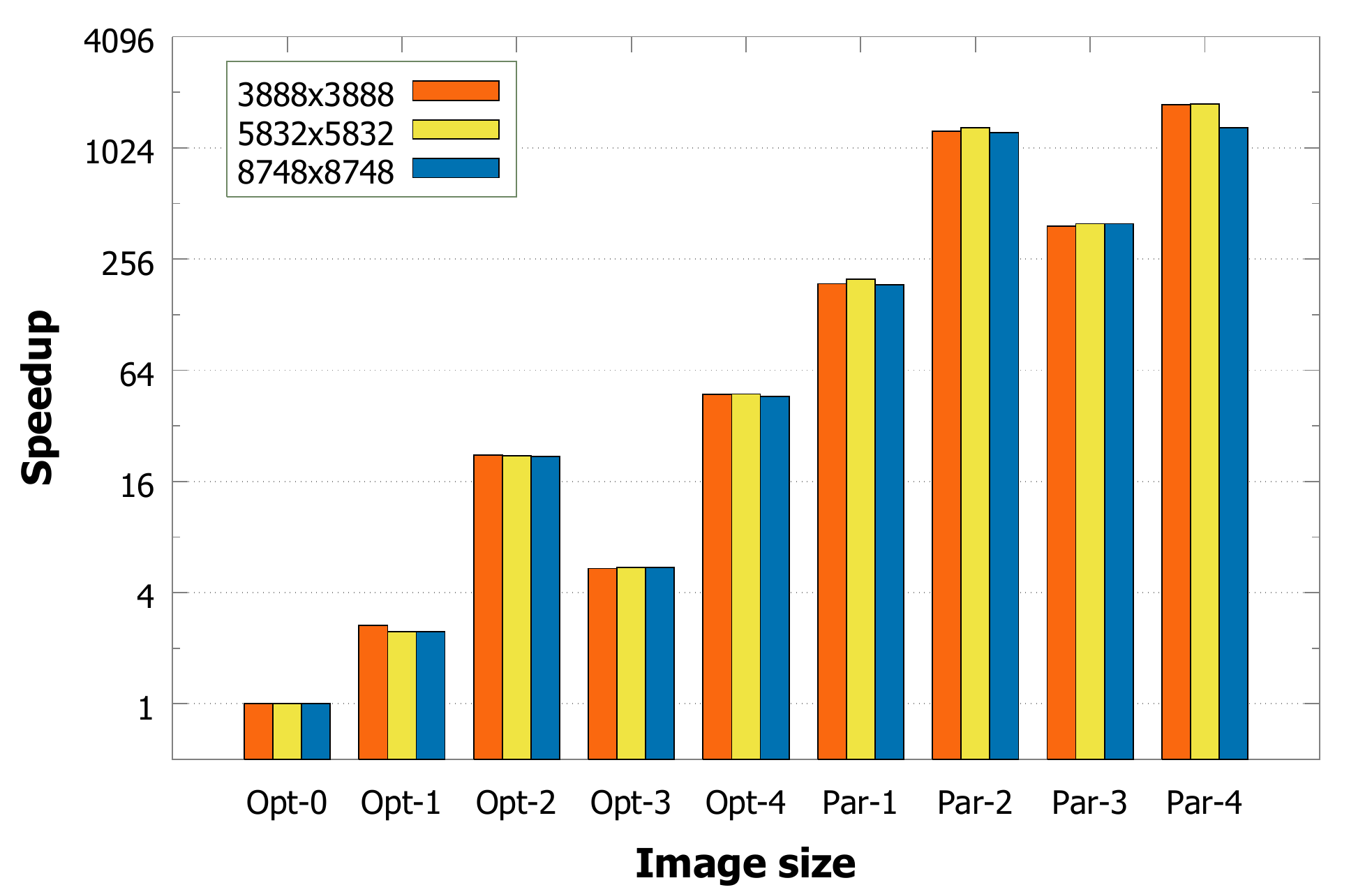}
	\caption{\textbf{From Naive to Parallelised Optimised code}
	\newline Baseline: single-pass algorithm \textbf{with} copy-back to source
	\newline Opt-0: Naive, Single-pass, No-vec
	\newline Opt-1: Single-pass, Unrolled, No-vec
	\newline Opt-2: Single-pass, Unrolled, SIMD
	\newline Opt-3: Two-pass, Unrolled, No-vec
	\newline Opt-4: Two-pass, Unrolled, SIMD	
	\newline Par-1 : Single-pass, Unrolled, No-vec, 100 omp threads
	\newline Par-2 : Single-pass, Unrolled, SIMD, 100 omp threads
	\newline Par-3 : Two-pass, Unrolled, No-vec, 100 omp threads
	\newline Par-4 : Two-pass, Unrolled, SIMD, 100 omp threads
	}
	\label{results:opt}
\end{centering}
\end{figure}

The speedup results for all the stages from naive to a parallelised optimised code are illustrated in Figure \ref{results:opt}.


\subsection{OpenMP Implementation Details}

An OpenMP implementation of the image convolution algorithm is shown in List. \ref{convAlg}. 

This code corresponds to the last stage of the optimisations, ``Par-4'', as it implements the two-pass algorithm with a horizontal pass followed by a vertical pass; the kernel loop is unrolled, \texttt{\#pragma simd} is used to enforce SIMD vectorisation, and the outer loop is parallelised.

It is worth stating that \texttt{\#pragma omp parallel for} has an implicit barrier at the end.


\begin{lstlisting}[caption={Two-pass Image Convolution Algorithm, OpenMP},label=convAlg]
/* 2D convolution on each plane */
void twoPassConv(float ***A, float ***B, float *k, int planeId, int rows, int cols) {
 // horizontal pass
#pragma omp parallel for
 for(int i=2; i<rows-2; i++) {
	#pragma simd
	 for(int j=2; j<cols-2; j++) {
	  B[planeId][i][j] =
             A[planeId][i][j-2] * k[0] +
             A[planeId][i][j-1] * k[1] +
             A[planeId][i][j] * k[2]   +
             A[planeId][i][j+1] * k[3] +
             A[planeId][i][j+2] * k[4];
	 }
 }
 // vertical pass
#pragma omp parallel for
 for(int i=2; i<rows-2; i++) {
	#pragma simd
	 for(int j=2; j<cols-2; j++) {
	  A[planeId][i][j] =
             B[planeId][i-2][j] * k[0] +
             B[planeId][i-1][j] * k[1] +
             B[planeId][i][j] * k[2]   +
             B[planeId][i+1][j] * k[3] +
             B[planeId][i+2][j] * k[4];
	 }
 }
return;}

/* calls twoPassConv on each plane */
void conv (float ***A, float ***B, float *ker, pimage a) {
#pragma novector
 for (int planeId= 0; planeId< a.planes; planeId++) {
	twoPassConv(A, B, ker, planeId, a.rows, a.cols);
 }
return;}
\end{lstlisting}

\subsection{OpenCL Implementation Details}

The concept of threads or tasks does not apply directly in OpenCL, but essentially OpenCL uses a model of \emph{compute units}, loosely corresponding to hardware threads, and \emph{processing elements}, which in a GPU maps to the ``cores'' in a streaming multiprocessor, but in the Xeon Phi maps to the vector units.  However, the most common programming model in OpenCL is to specify the ``global range", i.e. the number of invocations of a kernel, and let the runtime system allocate the threads. If one wants more fine-grained control, one can specify the ``local range" as well, which expresses the number of processing elements to use per compute unit.

Our approach for creating the OpenCL version from the original version of the code is largely automated.  We started from the ``Opt-3'' version discussed above. We use an annotation to mark the subroutine that will become the OpenCL kernel (i.e. \emph{twoPassConv}). Our source-to-source compiler\footnote{[omitted for blind review]} generates the OpenCL API code as well as a single-threaded OpenCL kernel. We then manually optimise the kernel and if required the OpenCL API calls. The OpenCL API we use is our own OpenCL wrapper library\footnote{[omitted for blind review]} , which provides convenient OpenCL integration in existing codebases for C, C++ and Fortran.

Deriving the parallel kernel from the generated single-threaded code is mostly a matter of replacing the loops by the OpenCL indexing calls (\texttt{get\_global\_id}, \texttt{get\_local\_id} etc), and in the case where the original code has multiple loops, as is the case for the convolution, we use a conditional statement to identifiy the portion of the kernel code to be executed on each invocation (List. \ref{convAlgOcl}). The host code contains a corresponding loop over the subsequent stages.

\begin{lstlisting}[caption={Two-pass Image Convolution Algorithm, OpenCL Kernel},label=convAlgOcl]
__kernel void twoPassConv(__global float *A,__global float *B,__global const float *k, __global const int* pass, const int cols, const int rows) {
    const int idx = get_global_id(0) ;
    const int c = idx % cols ;
    const int r = (idx  % (rows*cols)) / cols;
/* 2D convolution on each plane */    
    if (*pass==1) { 
    // horizontal pass    
    if(c>1 ) { if(c<cols-2 ) {
        A[idx] = B[idx-2] * k[0]
            + B[idx-1] * k[1]
            + B[idx] * k[2]
            + B[idx+1] * k[3]
            + B[idx+2] * k[4];
    }}} else if (*pass==2) { 
    // vertical pass
     if( r>1) { if(r<rows-2) {
        B[idx] =
            A[idx - 2*cols] * k[0] +
            A[idx - cols] * k[1] +
            A[idx] * k[2] +
            A[idx+cols] * k[3] +
            A[idx+ 2*cols] * k[4];
    }}} 
}
\end{lstlisting}

In order to validate our assumption about the mapping of work items to threads and vector units, we implemented the kernel in two different ways: first, the straightforward way, where only the global range is specified, and corresponds to the amount of work to be done,  i.e.\emph{npoints = rows*cols*planes} for the convolution.  The index into the image array is in this case simply the global index, because the OpenCL kernel uses a 1D representation of the array.

Then, the more controlled approach where the global range is \emph{ngroups*nths}, the local range is \emph{nths} and each kernel loops over \emph{niters = npoints/(ngroups*nths)}. Here, \emph{ngroups} is the number of work groups to be used,  usually this is the same as the number of compute units, and \emph{nths} is the number of work items per work group, usually the same as the number of processing  elements.
The index into the image array is given by \emph{idx = niters*nths*group\_id+nths*iter+local\_id}, the important point here is that the index is contiguous in the local id, rather than in the loop iterator. In this way the operations over \emph{nths} work items can be vectorised. 

We found that the optimal performance is achieved for \emph{ngroups =236} and \emph{nths=16}, which corresponds to 59 MIC cores with 4-way multithreading, i.e. 236 compute units, and 16 elements per SIMD vector (512-bit vectors). And indeed this is the same performance as achieved with the first, much simpler approach.

\subsection{GPRM Implementation Details}

The GPRM implementation of the two-pass algorithm defines the two phases of the algorithm as two different types of tasks. Since all the tasks defined in the GPC code will be executed in parallel, a \texttt{seq} pragma is required to run the two phases sequentially. Each phase uses a \emph{partial continuous for}, \texttt{par\_cont\_for} \cite{tousimojarad2014parallel}, in order to parallelise the outer loop over rows, and a \texttt{\#pragma simd} \footnote{Unlike OpenMP, in this case the use of \texttt{\#pragma simd} for the innermost loop in the GPRM implementation is optional} to help the compiler vectorise the inner loop over columns. \texttt{par\_cont\_for} is a sequential \emph{for} loop that works as follows: 

In GPRM, multiple instances  of the same task are generated (specified by \texttt{CUTOFF} in the List. \ref{convAlggprmTask}), each with a different index (similar to the \emph{global\_id} in
OpenCL). Each of these tasks calls \texttt{par\_cont\_for} passing its own index to specify which parts of the work should be performed by its host thread. 

\begin{lstlisting}[caption={Two-pass Image Convolution Algorithm, GPRM Task Code},label=convAlggprmTask]
#include "Conv.h"

void Conv::horizPass(ind, CUTOFF, ...) {
 par_cont_for(2, rows-2, ind, CUTOFF, this, &Conv::horizPassInnerLoop, ...);
}

void Conv::vertPass(ind, CUTOFF, ...) {
 par_cont_for(2, rows-2, ind, CUTOFF, this, &Conv::vertPassInnerLoop, ...);
}
\end{lstlisting}

\begin{lstlisting}[caption={Two-pass Image Convolution Algorithm, GPC Code},label=convAlggprmGPC]
#include "GPRM/Task/ConvTask.h"

void horizontalTasks (const int CUTOFF, ...) {
 /* GPC for with parallel evaluation */
 for (int ind=0; ind < CUTOFF ; ind++) {
   horizPass(ind, CUTOFF, ...);
 }
}

void verticalTasks ((const int CUTOFF, ...) {
 /* GPC for with parallel evaluation */
 for (int ind=0; ind < CUTOFF ; ind++) {
  vertPass(ind, CUTOFF, ...);
 }
}

void GPRM::ConvTask::twoPassConv(...) {
#pragma gprm seq
 {
   horizontalTasks(100, ...);
   verticalTasks(100, ...);
 }
}

\end{lstlisting}

\begin{table*}[ht]{
\centering
\renewcommand{\arraystretch}{1}
\caption{The effect of vectorisation on the parallel performance (ms) of the two-pass algorithm}
\tabcolsep=0.1cm
\hspace*{\fill} 
\begin{tabular}{c c c c c c c}
\toprule

Image Size & OpenMP no-vec & OpenCL no-vec & GPRM no-vec & OpenMP SIMD& OpenCL SIMD & GPRM SIMD\\
\midrule  \midrule

1152x1152 & 3.9 & 5.4 & 27.2 & 0.8 (4.9$\times$) & 2.0 (2.7$\times$) & 26.1 (1.0$\times$)\\
1728x1728 & 8.5 & 12.3 & 32.8 & 2.0 (4.2$\times$) & 3.8 (3.2$\times$) & 26.6 (1.2$\times$)\\
2592x2592 & 16.7 & 26.9 & 40.5 & 4.1 (4.1$\times$) & 7.8 (3.4$\times$) & 27.8 (1.5$\times$)\\
3888x3888 & 39.9 & 61.6 & 60.4 & 8.8 (4.5$\times$) & 16.5 (3.7$\times$) & 32.5 (1.9$\times$)\\
5832x5832 & 86.7 & 146.2 & 105.8 & 19.6 (4.4$\times$) & 38.1 (3.8$\times$) & 36.8 (2.9$\times$)\\
8748x8748 & 195.4 & 334 & 216.9 & 59.2 (3.3$\times$) & 91.5 (3.6$\times$) & 60.1 (3.6$\times$)\\

\hline 
\end{tabular}
\label{table:comparevector}
}
\hspace*{\fill} 

\end{table*}

\section{Parallel Performance of the Two-pass Algorithm}\label{conv:2-passResults}

The focus of this section is on the parallel performance of the three implementations of the two-pass algorithm.

We start by disabling the vectorisation in the Xeon Phi. The results for the parallelised non-vectorised cases are compared with the vectorised ones in Table \ref{table:comparevector}. In order to disable vectorisation for OpenMP and GPRM, the code should be compiled with the \texttt{-no-vec} flag.  In OpenCL, there is no explicit vectorisation option, but we can effectively disable vectorisation by using only a single processing element per compute unit.

The average speedup obtained through vectorisation for the OpenMP code is about 4.2$\times$. It is important to note that this speedup for the sequential code was almost twice as much (8.6$\times$). Therefore, the reported performance gain is specific to the case with 100 threads and should not be generalised.

It is worth noting that the speedup due to vectorisation in GPRM is much less pronounced, mostly due to the higher overhead of the GPRM runtime for smaller images. The same is true for OpenCL, but to a lesser extent. On average, the speedup obtained through vectorisation for the OpenCL code is about 3.5$\times$. Clearly, the OpenMP vectorisation is more efficient and this a large factor in the lesser performance of OpenCL.

\begin{figure}[h]
\begin{centering}
	\includegraphics[width=0.45\textwidth]{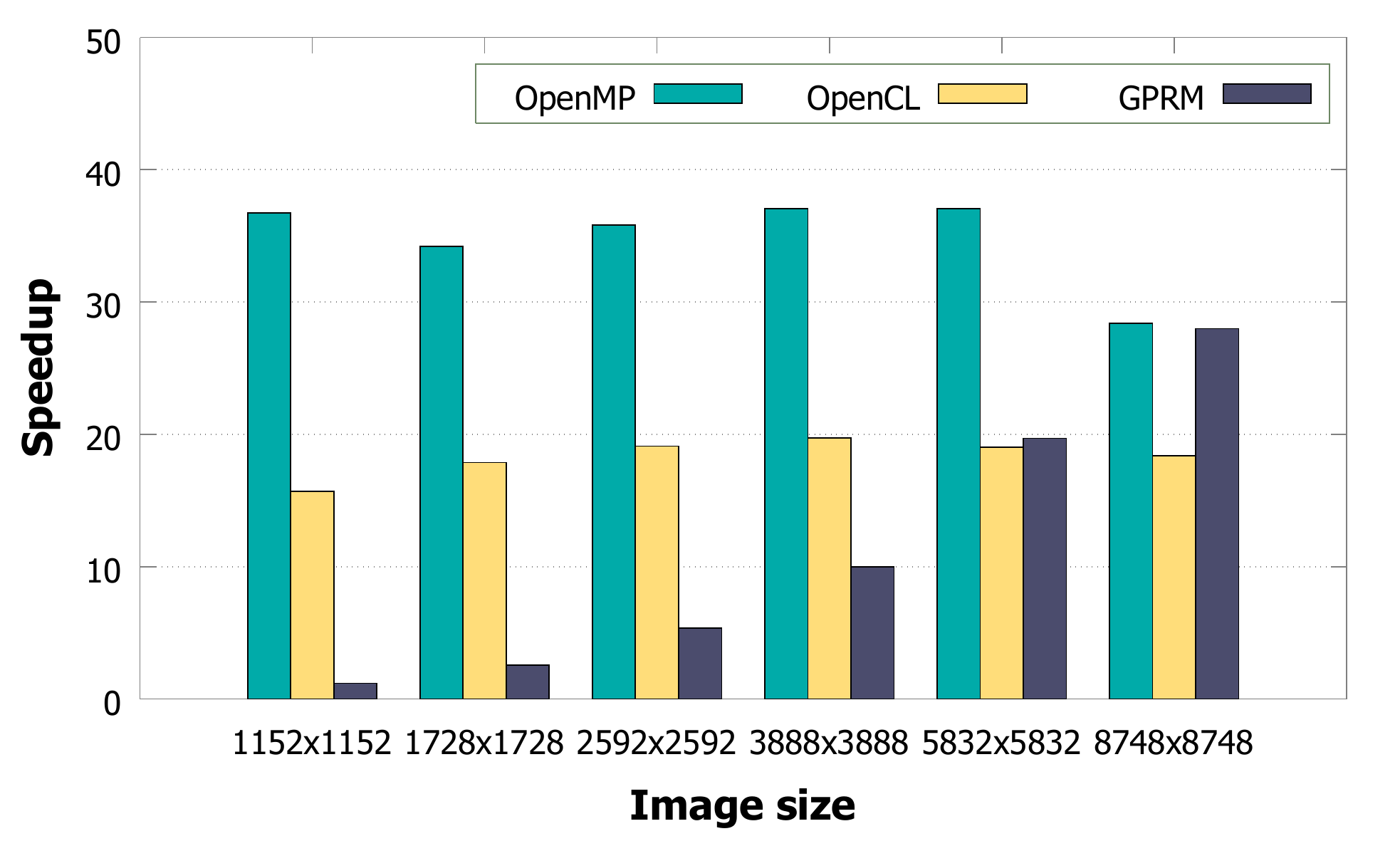}
	\caption{Speedup of the Vectorised Two-pass Algorithm, \textbf{R$\times$C}}
	\label{results:vectrc}
\end{centering}
\end{figure}

Figure \ref{results:vectrc} shows the speedup of the two-pass algorithm against its optimised sequential implementation (i.e. version ``Opt-4''). So far, the algorithm is parallelised over each plane of size R$\times$C, hence \textbf{R$\times$C} in Fig.\ref{results:vectrc}. This means for 3 colour planes, the parallelised code will be executed 3 times sequentially \footnote{Actually it is 3000 times, considering that we run the code 1000 times}.

It is possible to inspect the difference between OpenMP and GPRM more in detail. Since GPRM provides a modular methodology for expressing tasks and defining the communication patterns between them, we can create empty tasks and measure the overhead of distributing them across different threads and the parallel reduction. In other words, it is possible to measure the overhead of communication between tiles in GPRM. If we deduct this overhead from the total running time, we can measure the time spent on the actual computation inside the framework. The GPRM-compute time shown in Table \ref{table:overhead} is gained by deducting the constant communication overhead of 25.5ms from the total execution time. We will discuss how this separation will help to find a better task decomposition solution and hence better performance. Since the OpenMP execution model is different from that of GPRM, we cannot similarly separate the computation from the communication in OpenMP. OpenCL also allows to run empty kernels to study the overhead, we found that the overhead is between 0.25 and 0.4ms, so a  small component in the total running time for all but the smallest image size.

\begin{table}[ht]
{\centering
\renewcommand{\arraystretch}{1.2}
\caption{Running time (ms) per image for the two-pass algorithm}
\resizebox{1\columnwidth}{!}{
\tabcolsep=0.10cm
\scalebox{1} {
\begin{tabular}{ cccccc }
\toprule
Image Size&OpenMP&OpenCL&GPRM-total&OpenCL-compute&GPRM-compute\tabularnewline
\midrule  \midrule

1152x1152 &0.8& 2.0 &26.1&1.8&0.6\tabularnewline
1728x1728	&2.0& 3.8 &26.6&3.6&1.1\tabularnewline
2592x2592	&4.1& 7.8 &27.8&7.5&2.3\tabularnewline
3888x3888	&8.8& 16.5 &32.5&16.2&7.0\tabularnewline
5832x5832	&19.6& 38.1 &36.8&37.7&11.3\tabularnewline
8748x8748	&59.2& 91.5 &60.1&91.0&34.6\tabularnewline

\hline 
\end{tabular}
\label{table:overhead}
}
}
}
\end{table}

\begin{figure}[ht]
\begin{centering}
	\includegraphics[width=0.45\textwidth]{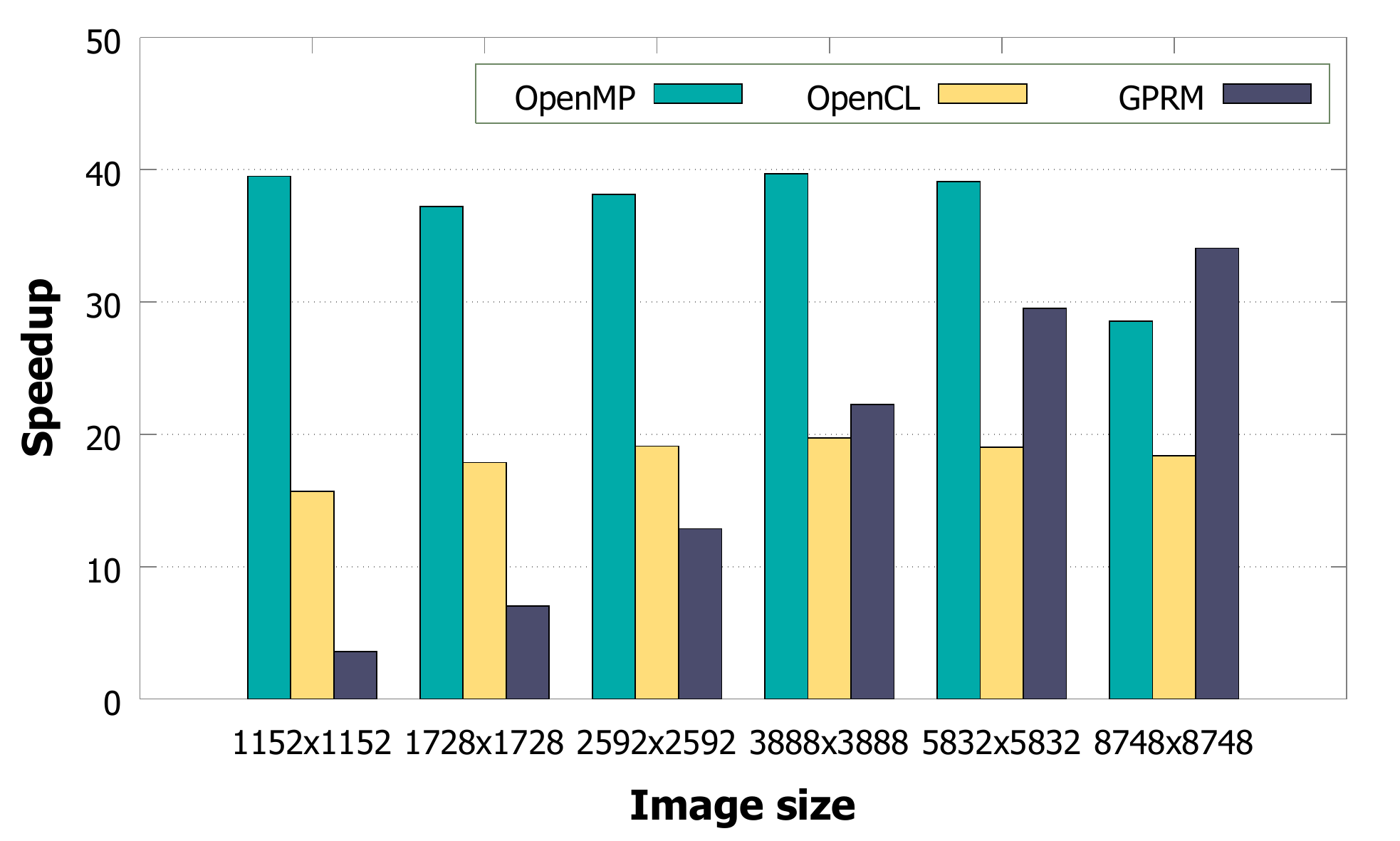}
	\caption{Speedup of the Vectorised Two-pass Algorithm, \textbf{3R$\times$C}}
	\label{results:vect3rc}
\end{centering}
\end{figure}

As a solution to mitigate the GPRM overhead, we have used task agglomeration: combining tasks into larger tasks to improve performance \cite{foster1995designing}. We therefore consider images with the width of 3 times the width of the original images, meaning that each row includes information for all 3 colour planes. This way, we include the 3 colour planes into the parallelisation. Using this technique, the size of tasks in GPRM becomes tripled and the overhead becomes one third (8.5ms per image). The speedup results for this case, which we call \textbf{3R$\times$C} is shown in Fig. \ref{results:vect3rc}. As expected, this technique does not have similar significant impact on the OpenMP performance. 

Using task agglomeration, GPRM achieves better performance than OpenCL for the three larger images, and the best performance among all for the largest image.

\section{Reconsidering the Single-pass Algorithm}\label{conv:nocpy1-passResults}

In order to compare the single-pass and two-pass algorithms, it is important to note that the two-pass algorithm uses an auxiliary array to store the result of the first pass. In the second pass, it uses the auxiliary array as the source and the original image as the destination, thus at the end of the algorithm, the original image will be replaced by the convolved one. It is convenient that the input and output images can use the same array, but it comes at a price: two assignment operations rather than one for every pixel. In order to have a fair comparison, we expected the same from the single-pass algorithm, i.e. overwriting the original image. This means that although the single-pass algorithm can produce the result on an output image by assigning new values for all the pixels only once, it now needs to copy the convolved values back to the original image. 

This copy-back operation constitutes a considerable extra overhead and sometimes is not needed, e.g. when working with the Xeon Phi as a co-processor. Suppose one runs some complex code on the Xeon CPU and offloads the computation of the convolution to the Xeon Phi, i.e. the typical model for OpenCL. In that model, there will be host-to-device and device-to-host copy cost. If one copies an image array A to the MIC, convolves it into an array B and copies that back to the host, there is of course no need to copy on the data back to the original array on the device itself. Consequently, we have also tested the single-pass code without the ultimate ``copy back to the original image'' operation. We have measured the results again only for the three larger images. 


\begin{figure}[h]
\begin{centering}
	\includegraphics[width=0.49\textwidth]{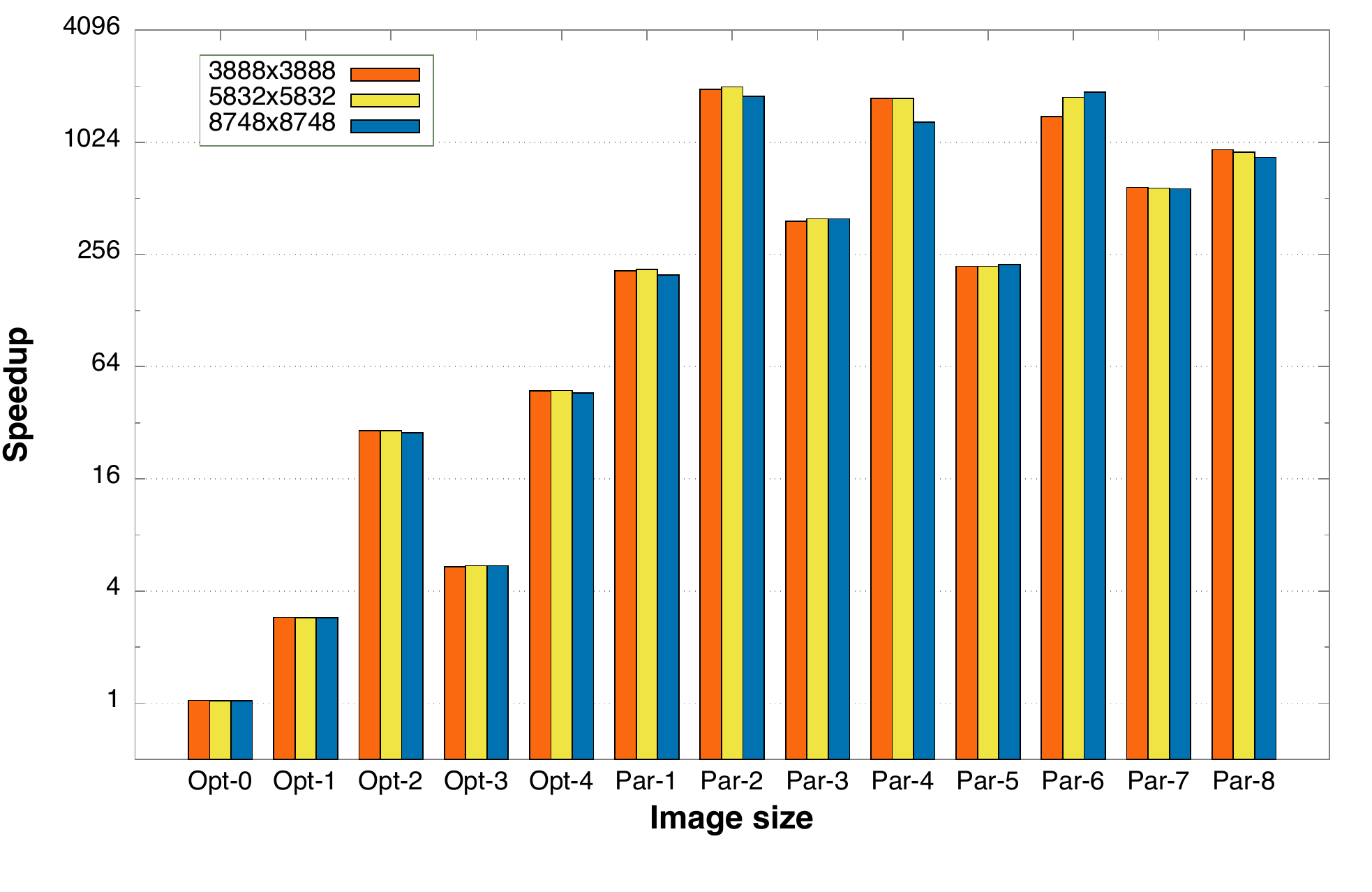}
	\caption{\textbf{From Naive to Parallelised Optimised code}
	\newline Baseline: single-pass algorithm \textbf{without} copy-back to source
	\newline Opt-0: Naive, Single-pass, No-vec
	\newline Opt-1: Single-pass, Unrolled, No-vec
	\newline Opt-2: Single-pass, Unrolled, SIMD
	\newline Opt-3: Two-pass, Unrolled, No-vec
	\newline Opt-4: Two-pass, Unrolled, SIMD	
	\newline Par-1 : Single-pass, Unrolled, No-vec, 100 omp threads
	\newline Par-2 : Single-pass, Unrolled, SIMD, 100 omp threads
	\newline Par-3 : Two-pass, Unrolled, No-vec, 100 omp threads
	\newline Par-4 : Two-pass, Unrolled, SIMD, 100 omp threads
	\newline Par-5 : Single-pass, Unrolled, No-vec, 100 GPRM tasks, 3R$\times$C
	\newline Par-6 : Single-pass, Unrolled, SIMD, 100 GPRM tasks, 3R$\times$C
	\newline Par-7 : Single-pass, Unrolled, SIMD, OpenCL
	\newline Par-8 : Two-pass, Unrolled, SIMD, OpenCL
	}
	\label{results:optnocpy}
\end{centering}
\end{figure}

After unrolling the kernel loop(s), for both non-vectorised and vectorised approaches, the results were as expected, i.e. the Two-pass algorithm had much better performance than the Single-pass algorithm (1.6$\times$ to 1.9$\times$). 

It is worth mentioning that in some cases, e.g. for 3888$\times$3888 images, the performance of the optimised single-pass algorithm with OpenMP could be improved by up to 15\% (10\% in average for the largest three images) by tuning the number of threads, e.g. with 120 threads, but since for the C++-based approaches 100 threads or tasks are used and we do not intend to compare multiple configurations together, we stick to this number. 

Figure \ref{results:optnocpy} shows that although the average sequential performance of the optimised two-pass code is 1.6$\times$ better than the average sequential performance of the optimised single-pass code (without copy-back), the average parallel performance of the optimised single-pass code (using OpenMP) is 1.2$\times$ better than that of the optimised two-pass code. The reason can also be extracted from Fig. \ref{results:optnocpy}: better utilisation of the vector units by the parallel single-pass code (9.4$\times$ its parallel non-vectorised version) compared to the parallel two-pass code (4.1$\times$ its parallel non-vectorised version). 

Since GPRM had shown a good performance for the largest array when we included parallelisation over planes into the tasks (the \textbf{3R$\times$C} case), its results has been added to Fig. \ref{results:optnocpy}. As expected, it produced the best result for the 8748$\times$8748 image, using the optimised single-pass algorithm with no copy-back. Its speedup over the baseline naive code is 1850$\times$ with 100 tasks.
 
The results of the OpenCL kernel for the single-pass implementation are on average about 50\% slower than for the two-pass implementation, and although the two-pass version still achieves over 1000$\times$ speedup over the baseline, it is still the worst of the three approaches.

As the best result amongst all, we have been able to get up to 1970$\times$ (for the 5832$\times$5832 image) speedup over the sequential naive implementation of the algorithm, by only utilising the compiler technology, few algorithmic changes, and parallelisation (using OpenMP). Also, a 2160$\times$ speedup over the baseline has been observed with 120 OpenMP threads for 5832$\times$5832 matrices with single-pass, no-copy approach.
 
\section{Related Work}
A similar 5$\times$5 spatial kernel (filter) has been the focus of a number of research papers \cite{petersen2013measuring} \cite{satish2015can} \cite{chimehcompiling} \cite{tian2015effective}. 

Petersen et al. \cite{petersen2013measuring} ported a subset of C benchmarks to Haskell and measured their performance on parallel machines, including the Xeon Phi. Considering three classes of naive, optimised, and Ninja C implementations \cite{satish2015can}, our implementation of the image convolution algorithm is classified as the optimised code, utilising loop unrolling and SIMD vectorisation. 

The reported Ninja gap for the Intel Labs Haskell Research Compiler (HRC) for 8192$\times$8192 images on the Xeon Phi using the single-pass algorithm is 3.7$\times$ (for 57 threads) \cite{petersen2013measuring}. The authors have disabled multithreading on the Xeon Phi, which is essentially different from hyper-threading on the Xeon processors \cite{jeffers2013intel}. 

Authors in \cite{chimehcompiling} explored this further and figured out that the peak performance can be achieved with 100 threads. They have also reported that the performance gap between the Vector Pascal \cite{chimehcompiling} and an optimised OpenMP implementations of the two-pass algorithm with 100 threads is almost 6.4$\times$. 

Authors in \cite{satish2015can} also focused on the optimisation techniques for parallel applications, using both advancements in compiler technology and algorithmic techniques to bring down the Ninja performance gap for throughput computing benchmarks, one of which is the single-pass implementation of the convolution algorithm.

Tian et al. \cite{tian2015effective} focused on efficient utilisation of the SIMD vector units on the Xeon Phi and proposed a number effective techniques to improve performance of parallel programs, including a single-pass image convolution. They have reported a speedup of 2000$\times$ using their vectorisation techniques along with parallelisation. We have also observed a speedup of about 1970$\times$ (2160$\times$ with 120 threads) without using any particular vectorisation technique. However, we have also highlighted the importance of the Xeon Phi vector units, specially their impact on parallel performance.

A similar study has been conducted using the TILEPro64 platform \cite{tousimojarad2016gprm}. On the 64-core TILEPro64, GPRM outperformed OpenMP in all cases.


\section{Discussion} 


The OpenCL implementation of the two-pass convolution performs worse than the OpenMP one by a factor of about two. This is actually not all that surprising because native OpenMP has very little overhead in its use of the kernel threads on the MIC, whereas OpenCL requires a runtime system for scheduling work on the threads. Furthermore, OpenCL has no explicit vectorisation control, so achieving good vectorisation is harder than with the pragmas used in the OpenMP code, as shown by our results.

The GPRM model also has a  fixed overhead (tens of milliseconds for hundreds of tasks) due to task creation and communication. By including the 3 image planes into the parallelisation (similar to the initial OpenCL approach), we reduced the overhead to one third and thus the GPRM  implementation achieved the best performance for the largest image. We therefore conclude that GPRM is not suitable to handle small tasks, but  as soon as the tasks become large enough to cover the scheduling overhead, it shows good performance compared to the competitive models.

GPRM naturally fits algorithms with task (functional) decomposition. It has its own complications though when it comes to domain decomposition, as it requires restructuring certain parts of the program to fit the GPRM structure. 

We therefore conclude that for the studied algorithm, 2D image convolution, OpenMP is the most productive approach, followed by OpenCL and then GPRM. In general, GPRM is attractive for task-based programming, but for algorithms like the convolution, it offers few advantages in terms of ease of programming. However, GPRM facilitates modular design, which is key to improve productivity \cite{hughes1989functional}. 

In terms of performance, OpenMP is the winning model, except for very large images where GPRM shows better performance after using task agglomeration. GPRM also outperformed OpenCL for the three largest images, but OpenCL had better performance for the three small images.

In terms of portability, GPRM and OpenMP only support conventional multicore platforms and the MIC, and require the commercial Intel compiler for the MIC. OpenCL has the advantage of supporting many platforms including GPUs and FPGAs, and its absolute performance is probably good enough for most purposes. Furthermore, using our described approach for porting applications, OpenCL programming becomes actually quite easy (as is clear from comparing the OpenCL kernel with the OpenMP code). It can also be noted that to program the Xeon Phi with OpenCL, one does not require the commercial Intel compiler, and the OpenCL SDK is free, so it is a cost-effective solution.


\section{Conclusion}
In this study, we have chosen three very different parallel programming models supported by the Xeon Phi (OpenMP, OpenCL and GPRM) to solve a 2D image convolution problem over a test set of 6 square images, ranging from 1152$\times$1152 to 8748$\times$8748. 
For a separable  convolution kernel, two different algorithms can be considered: Single-pass, which requires only a single assignment instead of two, but needs an additional copy if the result is required in the original array, and Two-pass, which requires fewer computations and returns the result in the original array. After creating optimised versions of both algorithms, we found that the choice between theses algorithm depends on which version of the single-pass algorithm is required: if the result has to be copied back to the original image, then the two-pass algorithm is always better. Otherwise, the single-pass algorithm can provide better parallel performance, even though its sequential performance is still worse.

We have explored a number optimisation and parallelisation techniques on the Xeon Phi which helped us achieve a speedup near 2000$\times$ over the baseline, but none of these techniques requires a major rewrite of the original code. The optimisation techniques include loop unrolling, vectorisation, and an algorithmic from single-pass to two-pass or vice versa. Task agglomeration is also used as a parallelisation technique to improve the performance of GPRM.


Although the OpenMP implementation showed the best overall performance on the Xeon Phi, the task-based GPRM model achieved better performance for large images, and although the OpenCL performance was not as good, it is still impressive, and thus a good choice if the Intel compiler is not available.




\bibliographystyle{IEEEtran}
\bibliography{ICPP2016}


\end{document}